\documentclass[aps,prb,twocolumn,showpacs,showkeys,groupedaddress]{revtex4-1}
\usepackage{graphicx}

\begin{document}


\title{Observation of interface carrier states in no-common-atom
heterostructures ZnSe/BeTe}


\author{A.S.~Gurevich}
\affiliation{A.F.Ioffe Physical-Technical Institute, 194021, St.
Petersburg, Russia }
\author{V.P.~Kochereshko}
\email[]{vladimir.kochereshko@mail.ioffe.ru}
\affiliation{A.F.Ioffe Physical-Technical Institute, 194021, St.
Petersburg, Russia }

\author{J.~Bleuse}
\affiliation{CEA-CNRS group ''Nanophysique et Semiconducteurs'', CEA, INAC/SP2M, and Institut N\'{e}el, 17 rue des Martyrs, F-38054 Grenoble, France  }

\author{H.~Mariette}
\affiliation{CEA-CNRS group ''Nanophysique et Semiconducteurs'', CEA, INAC/SP2M, and Institut N\'{e}el, 17 rue des Martyrs, F-38054 Grenoble, France  }

\author{A.Waag}
\affiliation{Braunschweig Technical University, Hans-Sommer-Stra$\beta $e 66, D-38106, Braunschweig, Germany }

\author{R.~Akimoto}
\affiliation{National Institute of Advanced Industrial Science and Technology (AIST), AIST Tsukuba Central 2-1, Tsukuba 305-8568, Japan}


\date{\today}

\begin{abstract}
Existence of intrinsic carrier interface states in heterostructures
with no common atom at the interface (such as ZnSe/BeTe) is
evidenced experimentally by ellipsometry and photoluminescence
spectroscopy. These states are located on interfaces and lie inside
the effective band gap of the structure; they are characterized by a
high density and high carrier capture rate. A tight binding model
confirms theoretically the existence of these states in ZnSe/BeTe
heterostructures for a ZnTe-type interface, in contrast to the case
of the BeSe-type interface for which they do not exist.
\end{abstract}


\maketitle

The main element of any semi-conductor nanostructure (quantum
wells, quantum wire, quantum dots) is an interface. Indeed,
there are no nanostructures without interfaces; additionally
the influence of interfaces grows with reduction of the
nanostructure's size. The imperfections of the interface limit
many basic parameters of nanostructures. For example, the
carrier mobility in HEMTs is limited by interface fluctuations,
the quantum efficiency of nanostructures is limited by
nonradiative recombination on interfaces, and the interfaces
are centers for segregation of impurities and defects, etc.

There have been many studies of extrinsic carrier states localized
on interfaces, for example states at the Si/SiO$_2$ interfaces in
MOS structures or interface dipole states in heterovalent structures
A$_2$B$_6$/A$_3$B$_5$. Besides such extrinsic interfaces states,
which can in principle be eliminated, the microscopic structure of
ideal interfaces in ideal heterostructures also can strongly
influence nanostructure properties.

In 1932 Tamm predicted the existence of carrier states localized at
the ideal surface of a semiconductor crystal\cite{Tamm1}. Such
states (Tamm surface states) are characterized by one-dimensional
localization along the direction normal to the surface and by energy
levels located inside the band gap of the bulk semiconductor. These
surface states, originating from clean and well ordered surfaces,
are intrinsic. They induce a Fermi level pining and lead to band
bending at the semiconductor surface. Since Tamm's prediction, many
theoretical and experimental studies have shown the strong influence
of such surface states on the optical and electrical properties of
semiconductors and semiconductor devices, see for example
\textcite{Dav2}.

James \cite{Dam3} was the first to suggest that similar states might
exist in the vicinity of a sharp interface separating two different
semiconductor materials. The analogy between Tamm and interface
states is based on the fact that a heterointerface, like the
surface, is a strong perturbation of the crystal's periodic
potential, which can lead to carrier localization.

Advances in molecular-beam epitaxy with growth control at the
monolayer  level have catalyzed scientific interest in research on
semiconductor interfaces. There has been much recent theoretical
work predicting the existence of the intrinsic interface states
\cite{Tok4}, but until now there have been practically no
experimental studies of such states, despite their obvious
importance for fundamental science and applications. Reference
\cite{Krom5} represented one of a few indirect experimental
observations of such states. Its authors studied the temperature
dependence of the lateral conductivity of quantum wells made of
materials with no common atom, namely InAs/AlSb. To explain this
dependence they postulate the existence of Tamm-like, localized
interface states not associated with defects but due to the strong
discontinuity in the periodic potentials on the two sides of the
interface.

\begin{figure}%
\includegraphics[width=0.8\linewidth]{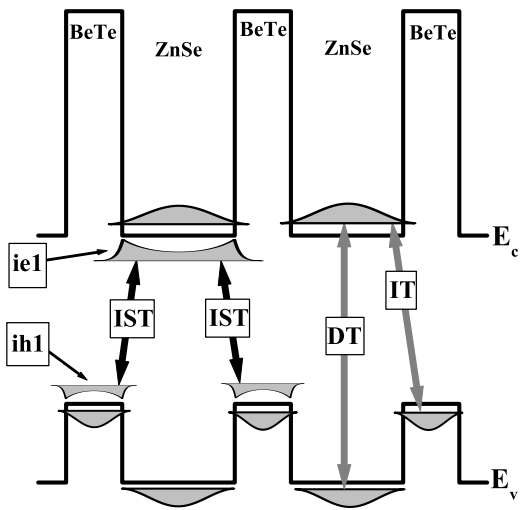}
\caption{%
Conduction and valence band of the studied structures, with their
electronic states and optical transitions. DT - direct exciton
transitions, IT - indirect exciton transitions, IST - carrier
recombination via electron (ie1) and hole (ih1) interface states. }
\label{figure1}
\end{figure}

In the structures with no-common atom on interface, for  example
AC/BD (here A and B are anions, C and D are cations), the interface
is a compound of two atoms: AD or CB which is different from the
main components of the heterostructure. The absence of a common atom
across the interface seems to lead to an additional
carrier-confining interface potential, thus promoting formation of
such interface states.

In ZnSe/BeTe quantum wells, another heterostructure with
"no-common-atom" interfaces, our investigations of the in-plane
optical anisotropy \cite{Gur6} gave some preliminary indications of
the possible existence of interface states in those structures.

In the present work we have performed a systematic study  of the
no-common-atom interface quantum well structure ZnSe/BeTe by low
temperature photoluminescence (PL) spectroscopy and spectrally
resolved ellipsometry. In association with a theoretical tight
binding study, the experiments have allowed us to answer the
following questions: (1) Do intrinsic electron (hole) states similar
to the Tamm surface state exist localized on the ideal interface?
(2) Are such states always present or are there some limitations on
their existence?

We also studied time resolved photoluminescence in these s
tructures. Our experimental data clearly demonstrate the existence
of electronic states strongly localized on heterostructure
interfaces.

Type-II ZnSe-based quantum well structures were grown on [100]  GaAs
substrates as a set of alternating ZnSe and BeTe layers with 20
periods in each sample. Some of the samples studied contained
manganese ions in the quantum well with a concentration of about 1\%
in the ZnSe layers, but we consider that Mn is not concerned in the
observation of interface states because of its low concentration and
because it is an isoelectronic atom which does not give rise to any
impurity levels. We studied about 10 samples with different well
widths. In all the samples, the width of every BeTe layer was equal
to half the width of the ZnSe layer and varied from 10 {\AA} to 100
{\AA}. The ZnSe and BeTe lattice constants almost coincide, i.e.
these layers do not experience any strain \cite{Nag7,Yam7}. From
previous work \cite{Gur6}, it was found that the band alignment
between ZnSe and BeTe layers is of Type II (see
figure~\ref{figure1}). This was deduced from the observation of an
indirect transition (IT) in the PL spectrum corresponding to a
spatially indirect recombination of electrons confined in ZnSe
layers and holes confined in BeTe layers. Since the barrier heights
in the conduction and valence bands are sufficiently large (Valence
Band Offset =0.9~eV, Conduction Band Offset =2.6~eV
\cite{Nag7,Yam7}), our samples can be considered as structures of
isolated quantum wells even at the minimum layer widths studied.

\begin{figure}%
\includegraphics[width=0.8\linewidth]{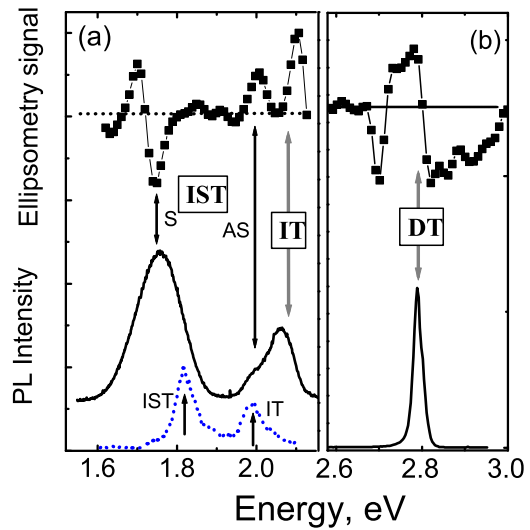}
\caption{%
(a) At bottom, solid line - Photoluminescence spectrum from
ZnSe/BeTe quantum well structure with 40 {\AA}/20 {\AA} layer
thicknesses, dotted line - Photoluminescence spectrum from ZnSe/BeTe
quantum well structure with 120 {\AA}/60 {\AA} layer thicknesses At
top - Ellipsometry spectrum from the same sample. Line IT
corresponds to the spatially indirect exciton
transitions, line IST   orresponds to the transitions via symmetric (S) and anti-symmetric (AS) interface states. \\
(b) is for a 100{\AA}/50{\AA} ZnSe/BeTe quantum well structure. At
bottom - photoluminescence, at top - ellipsometry. Line DT
corresponds to the spatially direct exciton in the ZnSe well.  }
\label{figure2}
\end{figure}

Two types of interfaces can be realized in such structures depending
on the growth conditions. They can be ZnTe-type interface or
BeSe-type interface, and they have been identified by the
observation of a strong optical anisotropy \cite{Yak8}. In our
structures it was demonstrated that all interfaces were of ZnTe type
\cite{Nag7}.

We performed a comparison of photoluminescence (PL)  spectra and
spectrally-resolved ellipsometry data. The ellipsometry spectra were
recorded in a wide spectral range and at temperatures varying from
10 to 77 K with a laboratory-built ellipsometer like that used in
reference\cite{Lee9}. Photoluminescence was excited by the third
harmonic (335 nm wavelength) of a YAG laser with diode pumping. Time
resolved spectra were obtained by exciting with the second harmonic
of a Ti:Sapphire laser and recording the emission with a streak
camera.

Figure 2 shows typical spectra of ZnSe/BeTe quantum  well
structures. Both ellipsometry and photoluminescence spectra exhibit
three main features: The relatively narrow line at high energy
(2.8~eV) labeled DT, which is observed in samples with wide ZnSe
layers, has been identified as due to the spatially direct excitonic
optical transition in the ZnSe layers. The very broad line labeled
IT at 2.05~eV corresponds to the indirect recombination of electrons
confined in ZnSe layers with holes in the BeTe layers.

In addition, a new line labeled IST (S), appears on the l ow energy
side in both luminescence and ellipsometry spectra in Fig.2. It is
very broad, with linewidth $\sim$ 100~meV. The IST emission is
absent in spectra of bulk ZnSe and BeTe. It can be observed in
heterostructures only at rather low excitation density. Its
intensity in both ellipsometry and PL spectra (which reflect
respectively the density of states and the carrier population)
indicates a large density of states and a high carrier-capture rate.
We connect this line with Interface State Transitions (IST).

The ratio of DT, IT, and IST line intensities depends on the  ZnSe
and BeTe layer widths, namely the intensity of the IST line relative
to the intensity of the IT line grows with decrease of the structure
period, while the intensity of the DT line falls.

The intensity of the IST PL line decreases when the excitation
density exceeds about 1~kW/cm$^2$, in contrast to the intensities of
the direct and indirect exciton PL lines (DT, IT) which increase
with increasing excitation density. The PL spectra of these lines
are compared in Fig.3 for two photoexcitation densities: 2~kW/cm$^2$
and 10~W/cm$^2$.

\begin{figure}%
\includegraphics[width=0.9\linewidth]{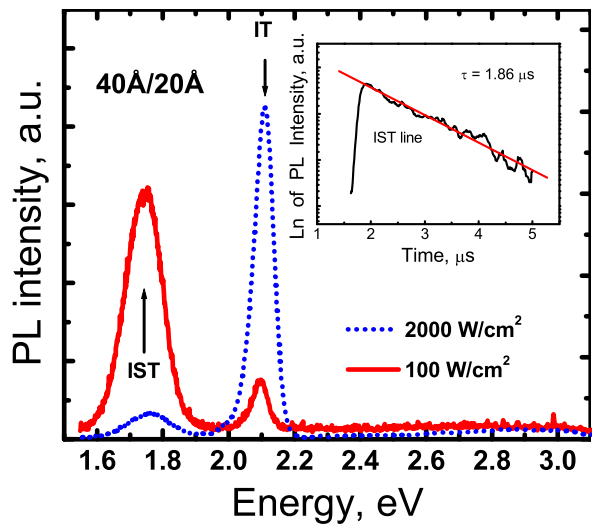}
\caption{%
Photoluminescence spectra taken from ZnSe/BeTe quantum well
structure with 40/20~{\AA}  layer thicknesses at two densities of
optical excitation 100~W/cm$^2$ and 2~kW/cm$^2$. Insert - PL decay
curve for the interface states transition (IST line). }
\label{figure3}
\end{figure}

At high excitation intensities, spatially separated electrons and
holes in the ZnSe and BeTe layers induce strong electric fields
inside the QW, which, in turn, give rise to a strong bending of the
conduction and valence bands. This leads to a strong change of the
electron and hole wave functions in the QWs and increase of their
overlap at the interface \cite{Zai10,Zai11,Brem15}. That leads to
growth of intensity of the IT line and to decrease of its radiative
decay time.

For the interface states the carrier wave functions are strongly
located on the interfaces and therefore these electric fields have a
weak influence on their overlapping and on the recombination time.

With decreasing of the lifetime the indirect recombination (IT)
becomes more effective, than carrier capture into the interface
states, and it results in redistribution of the intensity of these
lines.

We note that in previous work   \cite{Zai10} with similar
heterostructures, the IST line was not observed in PL spectra.
However the minimum excitation densities used in those experiments
were higher than our maximum excitation densities. At such high
excitation, the IST PL can no longer be observed.

When the temperature increases from 10~K to 77~K the intensity of all PL
lines decreases by a factor of $\propto $2. We note a high PL yield of the
studied structures at low temperatures, which indicates high quality of the
samples.

But the most important  peculiarity of the IST line is its
dependence on the heterostructure period. When the structure's
period decreases, the DT and IT lines shift towards higher energies,
while the IST (S) line shifts towards lower energies.

Figure~\ref{figure2} shows comparison of PL spectra for narrow
quantum wells  40/20~{\AA} (solid line) and wide wells 120/60~{\AA}
(dotted line). From this figure it is clear, that the PL line
related to the interface states (IST) is shifting to smaller energy
with narrowing of the well.

The shift to higher energy when  the period decreases is typical for
any confined state. By contrast the shift to lower energy for this
IST line is very unusual and requires an explanation.

It is well known that electron and hole states confined in a QW
increase  their energy with decrease of the confinement size
according to the uncertainty principle. Hence, the optical
transition (absorption or emission) should shift towards higher
energies as $1/L^2$, where $L$ is the well width.

Since the IST (S) line behaves in just the  opposite way, we can
conclude that the corresponding states are apparently not related to
quantization of the carriers in the wells.

This line shifts down in energy when the  barriers approach each
other. The known case for such behavior is realized in double
quantum wells, when the pair of states localized in the neighboring
wells start to interact and split into symmetric and anti-symmetric
states. In that case, the energy of the symmetric state decreases,
while the energy of the anti-symmetric state increases.

In order to explain the experimental findings  we propose a model
similar to that of double quantum wells. Every interface has a state
localized in the $Z$ direction with a localization size of several
angstroms. The energy level of this state is located in the gap
nearly 100~meV lower than the bottom of the QW. Thus, a quantum well
acts as a barrier separating these states on the neighboring
interfaces (Fig.1.). When the interfaces approach each other, these
states split into symmetric and anti-symmetric states. This leads to
a shift to lower energy of the symmetric state, and to a high energy
shift of the anti-symmetric state. In the PL spectra we see the
symmetric state(S) as a bright line because it is lower in energy. A
weak feature on the shoulder of the IT PL line in Fig.2 may
correspond to the antisymmetric state (AS). The anti-symmetric state
can be observed strongly in the ellipsometry spectrum, see
Fig.\ref{figure2}(a).

\begin{figure}%
\includegraphics[width=0.9\linewidth]{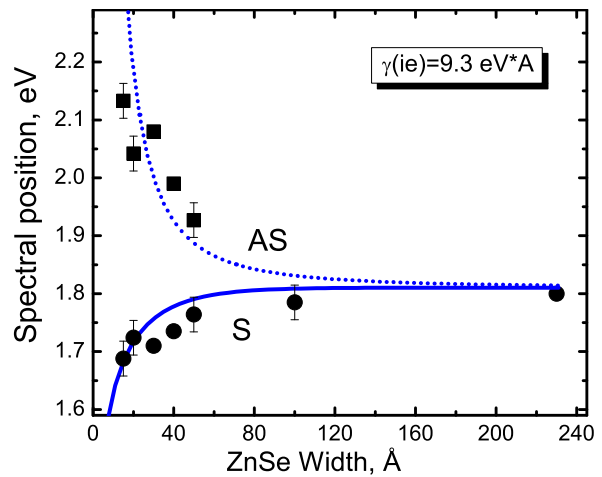}
\caption{%
Energy positions of the symmetric (S) and anti-symmetric (AS)
interface states as  a function of the ZnSe layer thickness in
ZnSe/BeTe quantum well structures. Symbols are experimental results,
squares -- ellipsometry, circles -- PL spectra. Curves --
calculation.} \label{figure4}
\end{figure}

The energy positions measured for the symmetric  and anti-symmetric
interface states and the results of the following calculation are
shown in Fig.~\ref{figure4}.

To calculate the localization energy for the  symmetric and
anti-symmetric states we solved a one dimensional Schr\"{o}dinger
equation with $\delta $ like potential on interfaces:
\[
-\frac{\hbar ^2}{2m}\frac{d^2\Psi }{dz}+U_{QW} (z)\Psi -\gamma [\delta
(z+a)+\delta (z-a)]\Psi =E\Psi
\]
Where: $U_{QW} (z)$ is a rectangular potential of the quantum well,
$2a$ is quantum  well width, $\gamma>0$ is a capacity of the
$\delta$ like potential. Fig.~\ref{figure4} shows the result of the
calculation with $\gamma=9.3$~eV*A .

As follows from these estimates \cite{Gur12}, the splitting of  the
hole states localized on the neighboring interfaces will be small,
due to a large hole mass, and will not explain the significant
magnitude of the symmetric state's energy shift observed in the
experiment. But the energy shift of the electron level, due to the
small effective mass of an electron, can reach as much as 100~meV.
Thus the energy shift of the IST line observed in the experiment is
mainly due to electron interface states. This conclusion, of course,
does not exclude the existence of the hole interface states, but
their energy shift with decrease of the structure period will be
significantly smaller than that of the electron states.

The last characteristic of this IST PL line is its decay time
behavior:  it is very long. Data were obtained by time resolved
spectroscopy as shown in the insert to Fig.3. The decay time for the
interface state transition is found to be as long as  2~$\mu $s,
that is three orders of magnitude longer than that of the indirect
exciton recombination (IT). At the same excitation density (about 1
kW/cm$^2$) the decay time of the IT line is about 5~ns. Such drastic
difference in the magnitude of the relaxation times for the IST line
compared to the other ones is a fingerprint of the very small
coherence volume of the electron-hole pair involved in this
transition \cite{Citrin1,Citrin2}, that is a strong localization of
the carriers on the interface. This also indicates that nonradiative
recombination through the IST is inefficient and there is no escape
of the carriers from the interface states, i.e. they are well
isolated from other states.

Moreover the decay of the IST PL is almost exponential and
independent of the excitation density, in contrast to the decay of
the IT PL which is non-exponential and very sensitive to excitation
power (from \textcite{Yak8} the decay time of the IT line varies
from tens of nanoseconds at weak excitation to hundreds of
picoseconds at high excitation as is expected for the indirect
recombination in type-II structures). Here, for the same range of
excitation conditions, the IST decay time shows no dependence on the
excitation intensity and on the structure period. This different
behavior of the IT and IST lines indicates that the carriers are
strongly localized at the same place (in the Z direction), in
contrast to the IT line for which the carriers are located in the
wells on both sides of the interface, and therefore are directly
affected by the electric field created by the excitation power.

The lack of dependence of the IST decay time on the structure
period indicates that the overlap of the electron and hole
wavefunctions of the interface states does not depend on the
structure period. That is, the IST states are strongly localized on
interfaces.

To summarize, it has been found experimentally that a new PL  line appears in multiple quantum well
ZnSe/BeTe structures, with the following features:\\
(1) It corresponds to a level located inside an effective bandgap at lower energy than the spatially
indirect exciton transitions. It is characterized by a high density of states and a high carrier capture rate.\\
(2) This line shifts toward low energy with decreasing thickness of the layers in the structures.\\
(3) The IST line disappears at high optical excitation.\\
(4) The decay time of the recombination through these IST states
does not depend on the excitation level and is several orders of
magnitude longer than the decay time through indirect exciton
recombination (IT) and many orders of magnitude longer than the
direct exciton recombination (DT).

All these results clearly indicate that this new line IST
corresponds to intrinsic carrier (electron and hole) states
localized on the interface, similar to the Tamm states on a crystal
surface.

To describe the interface carrier states a theoretical model  has
been developed \cite{Gur14} based on the empirical tight binding
model. In this model there are two important parameters:  $\beta$-
the overlap energy of the neighbor atoms on the interface and
$\Delta$ - the perturbation of the energy levels of the interface
atoms. From the numerical calculation, it follows that interface
states for electrons and holes can exist in a wide range of $\beta$
and $\Delta$ on the ZnTe-type interface. By contrast for a BeSe-type
interface, the interface states can not exist at any $\beta$ and
$\Delta$. A calculation for the case of GaAs/AlAs heterostructures
using the same tight binding model shows the absence of interface
states in this system. Indeed, a prerequisite condition for having
bound states at the interface is that the heterostructures should
have no common atom across the interface: in such cases the strongly
confining interface potential for both types of carriers leads to
the formation of interfaces states.

\section*{Conclusion} Studying ellipsometry and photoluminescence
spectra from ZnSe/BeTe quantum well structures, we found a new
spectral line which we attribute to interface states. This line, IST
(S), lies at lower energy than the effective bandgap in these
structures. It disappears from the PL spectra at high optical
excitation. With increasing confinement, contrary to all other
lines, it shifts to lower energies. We explain such a behavior as
follows: the states localized on neighbouring interfaces form
symmetric and anti-symmetric combinations, as in double quantum
wells. With decreasing structure period, the symmetric state shifts
down in energy while the anti-symmetric state shifts up. A
phenomenological model has been developed to describe the shift to
low energy. Calculations in the tight binding model show that the
interface states exist in structures with the ZnTe-type interface
and are absent on BeSe-type interfaces.

The interface between ZnSe and BeTe layers in our case is a giant
two dimensional ZnTe "molecule". The electron states of this
molecule are just our interface states. Because of monolayer
fluctuations of the QW width and barriers, this two dimensional
"molecule" splits into clusters with different energies of the
electron states that leads to the broadening of the IST and IT
lines.

Finally we have to emphasize that the interface states observed in
our work are intrinsic to this heterostructure and in principal
cannot be eliminated.

\begin{acknowledgments}  This work received supported from the RFBR,
the Presidium RAS, the Department of Physical Sciences RAS, and from
CNRS grants. We thank M.O. Nestoklon and A.V. Platonov for
discussions. Especially we thank Ronald T. Cox for help with
preparation of this paper.
\end{acknowledgments}

\bibliography{Interface_VK}
\end{document}